\documentclass[aps,prl,showpacs,twocolumn,superscriptaddress,amsmath,amsfont,amssymb,graphicx]{revtex4}
\usepackage{color,graphicx,epsfig} 
\usepackage{ifpdf}
\usepackage{amsmath} 
\usepackage{amssymb} 
\usepackage{bm}
\usepackage{color}
\usepackage{braket}

\usepackage[english]{babel}

\definecolor{Red}{rgb}{1.,0.,0.}

\begin{document}

\title{Flavor Changing Neutral Coupling  Mediated Radiative Top Quark Decays at Next-to-Leading Order in QCD}

\author{Jure Drobnak} 
\email[Electronic address:]{jure.drobnak@ijs.si} 
\affiliation{J. Stefan Institute, Jamova 39, P. O. Box 3000, 1001
  Ljubljana, Slovenia}

\author{Svjetlana Fajfer} 
\email[Electronic address:]{svjetlana.fajfer@ijs.si} 
\affiliation{J. Stefan Institute, Jamova 39, P. O. Box 3000, 1001
  Ljubljana, Slovenia}
\affiliation{Department of Physics,
  University of Ljubljana, Jadranska 19, 1000 Ljubljana, Slovenia}

\author{Jernej F. Kamenik}
\email[Electronic address:]{jernej.kamenik@ijs.si} 
\affiliation{J. Stefan Institute, Jamova 39, P. O. Box 3000, 1001
  Ljubljana, Slovenia}

\date{\today}

\begin{abstract}
We compute the branching ratios for the rare top quark decays $t\to c \gamma$ and $t \to c Z$ mediated by effective flavor changing neutral couplings at the next-to-leading order in QCD including the effects due to operator mixing. After re-suming contributions of order $[\alpha_s \log (\Lambda / m_t)]^n$, where $\Lambda$ is the scale at which the effective operators are generated, at leading log level using RGE methods, we compute finite matrix element corrections and study the effects of experimental kinematic cuts on the extracted branching ratios. We find that the $t \to c \gamma$ decay can be used to probe also the effective operators mediating $t \to c g$ processes, since these can naturaly contribute $10\%$ or more to the radiative decay. Conversely, any experimental signal of $t \to c g$ would indicate a natural lower bound on $t \to cZ, \gamma$.
\end{abstract}


\maketitle


The standard model (SM) predicts highly suppressed flavour changing neutral current (FCNC) processes of the top quark  ($t\to c V,\,V=Z,\gamma,g$) while new physics beyond the SM (NP) in many cases lifts this suppression (for a recent review c.f.~\cite{AguilarSaavedra:2004wm}). 

Top quark FCNCs can be probed both in production and in decays. Presently the most stringent bound on the $Br(t\to c Z)$ comes from a search performed by the CDF collaboration $Br(t\to c Z) < 3.7\%$ at $95\%$ C.L. \cite{:2008aaa}. For the photonic decay, the most stringent bound was put forward by the ZEUS collaboration $Br(t\to c \gamma) < 0.59 \%$ at $95\%$ C.L.~\cite{Chekanov:2003yt}. On the other hand the most stringent present limit on $Br(t\to c g)$ comes from the total cross-section measurement of CDF and yields $Br(t \to c g) < 0.57\%$ at $95\%$ C.L.~\cite{Aaltonen:2008qr}. The LHC will be producing about $80,000$ $t\bar t$ events per day at the luminosity $L = 10^{33} \mathrm{cm}^{-2}\mathrm{s}^{-1}$ and will be able to access rare top decay branching ratios at the $10^{-5}$ level with $10 \mathrm{fb}^{-1}$~\cite{Carvalho:2007yi}.

Recently~\cite{Zhang:2008yn} the $t\to c V$ decays mediated by effective FCNC couplings have been investigated at next-to-leading order (NLO) in QCD and it was found that $t\to c g$ receives almost $20\%$ enhancement while corrections to the $t\to c \gamma, Z$ branching ratios are much smaller.  However, the authors of~\cite{Zhang:2008yn} only considered a subset of all possible FCNC operators mediating $t\to c V$ decays at leading order and furthermore neglected the mixing of the operators induced by QCD corrections. 
In the case of $t\to c \gamma$ decay in particular the QCD corrections generate a nontrivial photon spectrum and the correct process under study is actually $t\to c \gamma g$. Experimental signal selection for this mode is usually based on kinematical cuts, significantly affecting the spectrum. The validity of theoretical estimates based on the completely inclusive total rate should thus be reexamined. Finally, renormalization effects induced by the running of the operators from the NP scale $\Lambda$ to the top quark scale are potentially  much larger than the finite matrix element corrections. Although these effects are not needed when bounding individual effective FCNC couplings from individual null measurements, they become instrumental for interpreting a possible positive signal and relating the effective description to concrete NP models.

In this Letter we present the results for the NLO QCD corrections to the complete set of FCNC operators mediating $t\to c V$ decays already at the leading order including operator mixing and renormalization effects. Finally in the case of $t\to c \gamma$ we study the effects of experimental kinematical cuts on the extracted branching ratio limits at NLO in QCD.


{\bf Basis of operators.} In writing the effective top FCNC Lagrangian we follow roughly the notation of ref.~\cite{AguilarSaavedra:2004wm, AguilarSaavedra:2008zc}. 
Hermitian conjugate and chirality flipped operators are implicitly contained in the Lagrangian and contributing to the relevant decay modes
\begin{equation}
\mathcal L^{tc}_{\mathrm {eff}} = \frac{v}{\Lambda^2} \sum_{V=g,\gamma,Z} b_{LR}^V \mathcal O_{LR}^V+ \frac{v^2}{\Lambda^2} a_L^Z \mathcal O_L^Z  + (L \leftrightarrow R) + \mathrm{h.c.}   \,,
\label{eq:L}
\end{equation}
where  $\mathcal O_{LR,RL}^V = g_V V^a_{\mu\nu} \bar q_{L,R} T^a \sigma^{\mu\nu} t_{R,L}$, $\mathcal O^Z_{L,R} = g_Z  Z_{\mu} \bar q_{L,R} \gamma^{\mu} t_{L,R}$, $q=c(u)$, $q_{R,L} = (1\pm\gamma_5)q/2$, $\sigma_{\mu\nu} = i[\gamma_\mu,\gamma_\nu]/2$, $g_Z = 2 e/\sin 2 \theta_W$, $g_\gamma = e$, $g_g = \sqrt{\alpha_s 4\pi}$. Furthermore $V(A,Z)_{\mu\nu} = \partial_{\mu} V_\nu - \partial_\nu V_\mu$, $G^a_{\mu\nu} =  \partial_{\mu} G^a_\nu - \partial_\nu G^a_\mu + g f_{abc} G_\mu^b G_\nu^c $, and $T^a$ are the Gell-Mann matrices in the case of the gluon and $1$ for the $\gamma, Z$. Finally $v=246$~GeV is the electroweak condensate and $\Lambda$ is the effective scale of NP.

Note that in principle, additional, four-fermion operators might be induced at the high scale which will also give (suppressed) contributions to $t\to c V$ processes. On the other hand, such contributions can be directly constrained via e.g. single top production measurements and we neglect their effects in the present study.


{\bf Operator renormalization.} The QCD virtual corrections to effective operators in eq. (\ref{eq:L}) involve ultra violet (UV) divergencies. These are cancelled exactly in the matching procedure to the underlying NP theory. The remaining logarithmic dependence on the matching scale can be resumed using renormalization group (RG) methods. 
The RG running is governed by the anomalous dimensions of the operators.
The operators $\mathcal O^Z_{L,R}$ do not mix with the others under QCD renormalization. They have identical anomalous dimensions of $\gamma^Z_{L,R} = \alpha_s C_F/\pi$, where $C_F = 4/3$. We assemble the remaining six operators into two vectors $\vec{\mathcal O_{i}} =  (\mathcal  \mathcal O^\gamma_i , \mathcal O^Z_i, O^g_i)^T$, $i=RL, LR$ which again do not mix with each other under QCD running. The corresponding one-loop anomalous dimension matrix is the same for both chiralities and reads
\begin{equation}
\gamma_i = \frac{\alpha_s}{2\pi}
\left[
\begin{array}{ccc}
 C_F  & 0   & 0   \\
0  & C_F  & 0   \\
 8 C_F / 3 &   C_F (3 - 8 s^2_W) / 3  & 5C_F - 2 C_A   
\end{array}
\right]\,,
\end{equation}
where $C_A= 3$ and $s_W^2=\sin^2\theta_W=0.231$ denotes the square of the sinus of the Weinberg angle. Depending on the nature of new physics which generates these dipole operators at the scale $\Lambda$, the relevant (LR) operators might explicitly include a factor of the top mass (i.e. by re-defining operators as ${\vec {\widetilde {\mathcal O}}_{LR}} =  (m_t/v )\vec  {\mathcal O}_{LR} $) and its running can then be taken into account by adding $6C_F$ to the diagonal entries of $\gamma_i$. As we shall demonstrate, this effect is numerically not important for the interesting range of couplings and scales, which can be probed at the Tevatron and the LHC. We are interested in particular in the mixing of the gluonic dipole contribution into the photonic and $Z$ dipole operators. For the case, without the top mass effects, the LR and RL operators receive identical corrections and the effective couplings at the top mass scale read 
\begin{subequations}
\begin{eqnarray}
 \hspace{-0.6cm}b^\gamma_{i} (\mu_t) \hspace{-0.15cm}&=& \hspace{-0.15cm} \eta ^{\kappa_1} b_i^\gamma (\Lambda )+\frac{16}{3}\left( \eta ^{\kappa_1}- \eta ^{\kappa_2}\right) b^g_i (\Lambda )\,,\\
 \hspace{-0.6cm}b^Z_{i} (\mu_t)  \hspace{-0.15cm}&=& \hspace{-0.15cm} \eta ^{\kappa_1} b_i^Z (\Lambda )   \hspace{-0.05cm}+\left[2-\frac{16}{3} s^2_W\right]\left( \eta ^{\kappa_1}- \eta ^{\kappa_2}\right) b^g_i (\Lambda )\,,
\end{eqnarray}
\end{subequations}
where $\mu_t$ is the top mass scale, $\eta = \alpha_s(\Lambda)/\alpha_s(\mu_t)$, $\kappa_1=4/3\beta_0$, $\kappa_2=2/3\beta_0$ and $\beta_0$ is the one-loop QCD beta function. Assuming that no new colored degrees of freedom appear below the UV matching scale which would modify the QCD beta function, it evaluates to $\beta_0=7$ above the top mass scale. If we include the top mass running in the RGE of LR operators, then $\kappa_{1,2}$ are modified to $\kappa_1=16/3\beta_0$, $\kappa_2=14/3\beta_0$. 

\begin{figure}[t]
\begin{center}
\includegraphics[width=8cm]{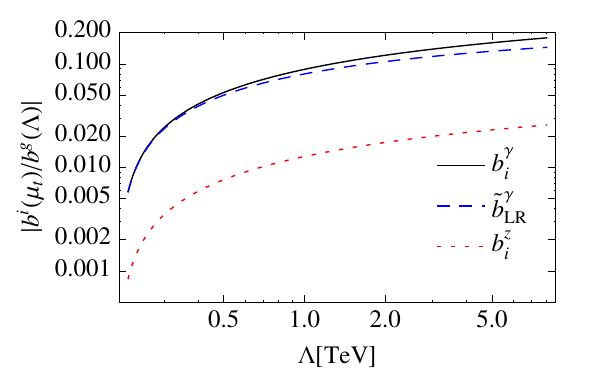}
\end{center}
\caption{ \label{fig:1} Effective $tc\gamma$ and $tcZ$ couplings at the top mass scale $\mu_t\approx 200$GeV induced by the RGE running of the $tcg$ coupling from the NP scale $\Lambda$.}
\end{figure}

We illustrate the effect of the RGE running in Figure \ref{fig:1} where we plot the values of $b_i^{\gamma,Z}$ at the top mass scale $\mu_t\simeq 200$ GeV induced solely by the mixing of the gluonic dipole contribution, produced at the UV scale $\Lambda$.
We see that for NP matching scales above $2$ TeV the induced contributions to $b^\gamma_{LR,RL}$ are around 10\% of the $b^g_{LR,RL}$ in the UV. On the other hand, due to cancelations in the RGE equations for the $b^Z_{LR,RL}$, these receive much smaller corrections (below 1 \% for the interesting range). Including the top mass renormalization reduces the induced corrections to the $\widetilde b^\gamma_{LR}$ coupling. However for UV scales of a couple of TeV or below, this effect is negligible.


{\bf Matrix element corrections.} To consistently describe rare top decays at NLO in $\alpha_s$ one has to take into account finite QCD loop corrections to the matrix elements $\bra{q \gamma} \mathcal O_i \ket{t}$ and  $\bra{q Z} \mathcal O_i \ket{t}$ evaluated at the top mass scale as well as single gluon bremsstrahlung corrections, which cancel the associated infrared and collinear divergencies in the decay rates. Contributions due to the $O^{\gamma,Z}_{LR,RL}$ have already been computed in ref.\cite{Zhang:2008yn}. Here we present results for the $\mathcal O^{Z}_{L,R}$ as well as for the admixture of the gluonic dipole operators $\mathcal O^g$. The $t\to c V$ rate for $V=Z,\gamma$ with implicit operator chirality assignments ($a$ and $b$ stand for $a_L$ and $b_{LR}$ or $a_R$ and $b_{RL}$) reads
\begin{eqnarray}
\Gamma^V \hspace{-0.2cm}&=& \hspace{-0.15cm} \frac{v^3m_t}{\Lambda^4} \left[2\mathrm{Re}\{a^{V*}b^g\} \Gamma^V_{ag} +2\mathrm{Re}\{b^{V*}a^V\} \Gamma^V_{ab} \right.\\
 &&\hspace{-0.15cm} \left.  - 2\mathrm{Im}\{a^{V*}b^g\} \tilde{\Gamma}^V_{ag}\right]   + \frac{v^2 m_t^2}{\Lambda^4}\left[ |b^V|^2 \Gamma^V_{b} 
+|b^g|^2 \Gamma^V_{g} \right.\nonumber\\
&&\hspace{-0.15cm}\left.+2\mathrm{Re}\{b^{V*}b^g\} \Gamma^V_{bg} -2\mathrm{Im}\{b^{V*}b^g\}\tilde{\Gamma}^V_{bg} \right] + 
 |a^V|^2\frac{v^4}{\Lambda^4}      \Gamma_{a}^V  \,,\nonumber
\end{eqnarray}
where $a^\gamma=0$.  At the tree level only $\Gamma^V_{a,b,ab}$ contribute and without imposing any kinematical cuts we derive them in $d=4+\epsilon$ dimensions and for massless charm quark as
\begin{subequations}
\begin{eqnarray}
\Gamma_{b}^{V(0)}\hspace{-0.20cm}&=&\hspace{-0.20cm}\frac{m_t}{16\pi}g_V^2(1-r_V)^2 \Gamma(1+\frac{\epsilon}{2})(1-r_V)^{\epsilon} 2(2+\epsilon+r_V)\,,\nonumber\\
&&\\
\Gamma_{a}^{Z(0)}\hspace{-0.2cm}&=&\hspace{-0.20cm}\frac{m_t}{16\pi}g_Z^2(1-r_Z)^2 \Gamma(1+\frac{\epsilon}{2})(1-r_Z)^{\epsilon}\nonumber\\
&&\times \frac{1}{2r_Z}\big[1+(2+\epsilon)r_Z\big]\,,\\
\Gamma_{ab}^{Z(0)}\hspace{-0.2cm}&=&\hspace{-0.20cm}\frac{m_t}{16\pi}g_Z^2(1-r_Z)^2 \Gamma(1+\frac{\epsilon}{2})(1-r_Z)^{\epsilon}(3+\epsilon)\,,
\end{eqnarray}
\end{subequations}
where $r_Z = (m_Z/m_t)^2$ and $r_\gamma=0$. We first study $\alpha_s$ virtual and real bremsstrahlung corrections to these contributions leaving $\mathcal O^g$ operator contributions aside.  We obtain
\begin{eqnarray}
\Gamma_{b}^V &=& \Gamma_{b}^{V(0)} \left\{1+\frac{\alpha_s}{4\pi}C_F
\Bigg[
2\log\left(\frac{m_t^2}{\mu^2}\right) 
+ 8 
-8\mathrm{Li}_2(r_V)
\right.\nonumber\\
&&\hspace{-1.3cm}
-\frac{4\pi^2}{3}
- 4 \log(1-r_V)\log(r_V)
-\frac{2(8+r_V)}{2+r_V}\log(1-r_V)
 \nonumber\\
&&\hspace{-1.3cm}\left.
- \frac{4r_V(2-2r_V-r_V^2)}{(1-r_V)^2(2+r_V)}\log(r_V)
-\frac{16-11r_V-17r_V^2}{3(1-r_V)(2+r_V)} 
\Bigg]\right\},\label{eq:Gb}\nonumber\\
&&\\
\Gamma^Z_{a} &=&\Gamma_{a}^{Z(0)}\left\{1+\frac{\alpha_s}{4\pi}C_F
\Bigg[
-4\log(1-r_Z)\log(r_Z) 
\right.\nonumber\\
&&\hspace{-1.3cm}
-2\frac{5+4r_Z}{1+2r_Z}\log(1-r_Z)
-\frac{4r_Z(1+r_Z)(1-2r_Z)}{(1-r_Z)^2(1+2r_Z)}\log(r_Z)
\nonumber\\
&&\hspace{0.1cm}\left.
+\frac{5+9r_Z-6r_Z^2}{(1-r_Z)(1+2 r_Z)}
-8\mathrm{Li}_2(r_Z) 
-\frac{4\pi^2}{3}
\Bigg]\right\}\,,\\
\Gamma^Z_{ab} &=& \Gamma_{ab}^{Z(0)}\left\{ 1 + \frac{\alpha_s}{4\pi}C_F
\Bigg[
\log\left(\frac{m_t^2}{\mu^2}\right)
+4
-8\mathrm{Li}_2(r_Z)
\right.\nonumber\\
&&\hspace{-1.3cm}
-\frac{4\pi^2}{3}
-4\log(1-r_Z)\log(r_Z) 
-\frac{2(2+7r_Z)}{3r_Z}\log(1-r_Z)
\nonumber\\
&&\hspace{0.5cm}\left.
-\frac{4r_Z(3-2r_Z)}{3(1-r_Z)^2}\log(r_Z)
+ \frac{5-9r_Z}{3(1-r_Z)}
\Bigg]\right\}\,.
\end{eqnarray}
The $\alpha_s$ corrections to $\Gamma_b^i$ in eq. (\ref{eq:Gb}) have been derived before \cite{Zhang:2008yn} while the remaining two expressions are a new result. We confirm the result for $\Gamma_b^\gamma$, which can most easily be obtained from the corresponding $B\to X_s \gamma$ expressions, which have been known for some time \cite{Ali:1995bi}. On the other hand we find disagreement with the nonzero $r_Z$ dependence reported in \cite{Zhang:2008yn}. We have crosschecked our result with the corresponding calculation done for a virtual photon contributing to $B \to X_s \ell^+ \ell^-$ \cite{Asatryan:2001zw}. In the case of $\Gamma_a^Z$ our result agrees with the corresponding calculation of $t\to b W$ at NLO \cite{tbW}. In fact, when considering the branching ratio $Br(t\to c Z)$ approximated as $\Gamma(t\to c Z ) / \Gamma(t \to b W)$, $\alpha_s$ corrections  cancel almost exactly for $\Gamma_{a,b}^Z$ contributions leaving corrections of  $  (\alpha_s / 4\pi) 6 C_F (r_W-r_Z)  [1 + \mathcal O(r_W,r_Z)] = -0.0006$ and $ (\alpha_s / 4\pi) (C_F / 3 )[1 + \mathcal O(r_W,r_Z)] = 0.001$ respectively~\footnote{The numerical values are computed using the full $r_i$ dependence and $\alpha_s(m_t)=0.107$}. In particular, the cancelation in $b^Z$ contributions is even more severe than reported in  \cite{Zhang:2008yn}.

\begin{figure}[t]
\begin{center}
\includegraphics[width=5.5cm]{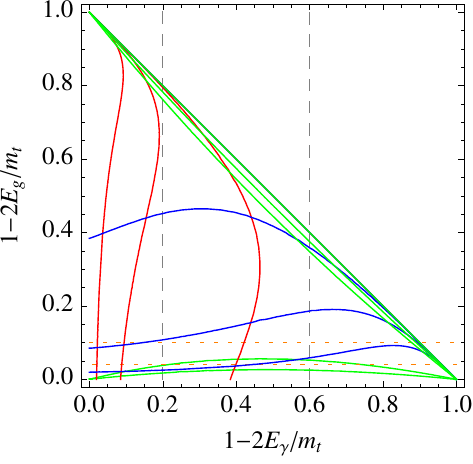}
\end{center}
\caption{ \label{fig:dalitz} The $t\to c \gamma g$ Dalitz plot. Contours of constant photon and gluon infrared and collinear divergent contributions are drawn in red and blue respectively. The collinear divergencies appear at the horizontal and vertical boundaries of the phase-space, while the IR divergencies sit in the top and right corners. The cuts on the photon energy correspond to vertical lines, the cuts on the gluonic jet energy to horizontal lines. Green lines correspond to cuts on jet veto cone around the photon. }
\end{figure}

\begin{table}
\begin{center}
\caption{\label{table:num}Numerical values for coefficient functions corresponding to the inputs
$m_t = 172.3\,\mathrm{GeV}\,,$  $m_Z = 91.2\,\mathrm{GeV}\,,$  $\sin^2\theta_W = 0.231\,.$}
\begin{tabular}{llll}
\hline\hline
$x_{b}=2.36$ & $x_{a}=1.44$ & $x_{ab}=1.55$   \\
$y_{b}=-17.90$ & $y_{a}=-10.68$ & $y_{ab}=-10.52$ &$y_{g}=0.0103$\\
 $y_{bg}=3.41$ & $y_{ag}=2.80$ & $\tilde{y}_{bg}=2.29$ &  $\tilde{y}_{ag} = 1.50$\\
 \hline\hline
\end{tabular}
\end{center}
\end{table}
Including contributions from gluonic dipole operators the analytic expressions are rather lengthy and will be presented elsewhere \cite{long}. For the case of the $Z$ boson we find for the completely inclusive rate
\begin{eqnarray}
\Gamma&=&\frac{m_t}{16\pi}g_Z^2\Bigg\{
|a^Z|^2 \frac{v^4}{\Lambda^4} \Big[x_{a} +\frac{\alpha_s}{4\pi}C_F y_{a} \Big] \\
&&\hspace{-0.5cm}+2\frac{v^3 m_t}{\Lambda^4} \left[ \mathrm{Re}\{b^{Z*} a^Z\}\Big(x_{ab} +\frac{\alpha_s}{4\pi}C_F y_{ab} \Big)\right.\nonumber\\
&&\hspace{-0.5cm}\left. +\mathrm{Re}\{a^{Z*}b^g\}\frac{\alpha_s}{4\pi}C_F y_{ag}-\mathrm{Im}\{a^{Z*}b^g\}\frac{\alpha_s}{4\pi}C_F \tilde{y}_{ag}\right]\nonumber\\
&&\hspace{-0.5cm}+\frac{v^2m_t^2}{\Lambda^4} \left[ |b^Z|^2 \left(x_{b}+\frac{\alpha_s}{4\pi}C_F y_{b}  \right) + |b^g|^2\frac{\alpha_s}{4\pi}C_F y_{g}\right.\nonumber\\
&&\hspace{-0.5cm}\left.+2\mathrm{Re}\{b^{Z*}b^g\}\frac{\alpha_s}{4\pi}C_F y_{bg}-2\mathrm{Im}\{b^{Z*}b^g\}\frac{\alpha_s}{4\pi}C_F \tilde{y}_{bg} \right] \Bigg\}\nonumber\,,
\end{eqnarray}
where the numerical $x_i,y_i,\tilde y_i$  coefficients are given in Table \ref{table:num}. The resulting $\alpha_s$ corrections to $Br(t\to c Z)$ are at the order of a few percent for $|b^g| \sim |b^Z|, |a^Z|$. In the case of $t\to c\gamma$ at NLO the process in question is $t\to c \gamma g$ and involves three (one almost) massless particles in the final state. Virtual matrix element corrections contribute only at the soft gluon endpoint ($E_g = 0$) and result in non-vanishing  $b^\gamma b^g$ interference contributions. They involve IR divergencies which are cancelled by the real gluon emission contributions. These also produce non-vanishing $|b^g|^2$ contributions, and  create a non-trivial photon spectrum involving both soft and collinear divergences. The later appear whenever a photon or a gluon is emitted collinear to the light charm jet.  In the analogous $B\to X_s\gamma$ decay measured at the $B$-factories the photon energy in the $B$ meson frame can be reconstructed and a hard cut ($E_\gamma^{cut}$) on it removes the soft photon divergence. The cut also ensures that the $B\to X_s g$ process contributing at the end-point $E_\gamma = 0$ is suppressed. On the other hand, in present calculations the collinear divergencies are simply regulated by a non-zero strange quark mass, resulting in moderate $\log(m_s/m_b)$ contributions to the rate.  The situation at the Tevatron and the LHC is considerably different.  The initial top quark boost is not known and the reconstruction of the decay is based on triggering on isolated hard photons with a very loose cut on the photon energy (a typical value being $E_\gamma>10 $\,GeV in the lab frame \cite{Aad:2009wy}). Isolation criteria are usually specified in terms of a jet veto cone $\Delta R = \sqrt {\Delta \eta^2 + \Delta \phi^2}$ where $\Delta\eta$ is the difference in pseudorapidity and $\Delta\phi$ the difference in azymuthal angle between the photon and nearest charged track. Typical values are $\Delta R > (0.2-0.4)$ \cite{CDF_2007_06}. We model the non-trivial cut in the top quark frame with a cut on the projection of the photon direction onto the direction of any of the two jets ($\delta r = 1- {\bf p_\gamma \cdot p_j} / E_\gamma E_j$). The effects of the different cuts on the Dalitz plot are shown in Figure \ref{fig:dalitz}.
Since at this order there are no photon collinear divergencies associated with the gluon jet, the $\delta r$ cut around the gluon jet has numerically negligible effect on the rates. On the other hand the corresponding cut on the charm jet - photon separation does not completely remove the collinear divergencies in the spectrum. However, they become integrable. The combined effect is that the contribution due to the gluonic dipole operator can be much more pronounced than in the case of $B\to X_s \gamma$. 
\begin{figure}[t]
\begin{center}
\includegraphics[width=5.5cm]{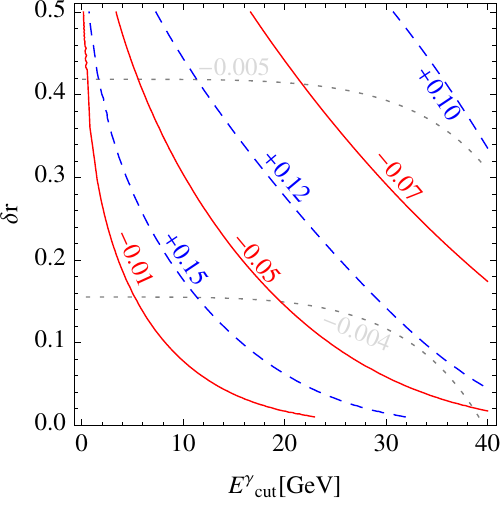}
\end{center}
\caption{ \label{fig:drEcut}  Relative size of $\alpha_s$ corrections to the $Br(t\to c \gamma)$ at representative ranges of $\delta r$ and $E_{cut}^\gamma$. Contours of constant correction values are plotted for $b^g=0$ (gray, dotted), $b^g=b^\gamma$ (red) and $b^g = - b^\gamma$ (blue, dashed).}
\end{figure}

The analytic formulae for the $\Gamma^\gamma_{bg,g}$ and $\tilde \Gamma^\gamma_{bg}$ with full $\delta r$ and $E^{cut}_\gamma$ dependence are rather lengthy and will be presented elsewhere \cite{long}.  In Figure \ref{fig:drEcut} we show the $b^g$ induced correction to the tree-level $Br(t\to c \gamma)$ for representative ranges of $\delta r$ and $E^\gamma_{cut}$.
We observe, that the contribution of $b^g$ can be of the order of $10-15\%$ of the total measured rate, depending on the relative sizes and phases of $\mathcal O_{LR,RL}^{g,\gamma}$ and on the particular experimental cuts employed. Consequently, a bound on $Br(t\to c \gamma)$ can, depending on the experimental cuts, probe both $b^{g,\gamma}$ couplings.

In summary, QCD corrections to FCNC coupling mediated rare top decays can induce sizable mixing of the relevant operators, both through their renormalization scale running as well in the form of finite matrix element corrections. These effects are found to be relatively small for $t\to c Z$ decays. On the other hand the accurate interpretation of experimental bounds on radiative top processes in terms of effective FCNC operators requires the knowledge of the experimental cuts involved and can be used to probe $\mathcal O^g_{LR,RL}$ contributions indirectly.







\begin{acknowledgments}
 J. F. K. would like to thank Mikolaj Misiak for useful discussions and the Galilo Galilei Institute for Theoretical Physics for the hospitality and the INFN for partial support during the completion of this
work. This work is supported in part by the European Commission RTN
  network, Contract No. MRTN-CT-2006-035482 (FLAVIAnet) and by the Slovenian Research
  Agency.
\end{acknowledgments}

\appendix

\end{document}